\documentclass[11pt]{article}

\usepackage{epsfig}
\usepackage{graphicx}
\usepackage{amsmath}
\usepackage{rotating}
\usepackage{authblk}

\newcommand{\be}{\begin{eqnarray}}
\newcommand{\ee}{\end{eqnarray}}
\newcommand{\ba}{\begin{array}}
\newcommand{\ea}{\end{array}}
\newcommand{\bt}{\begin{tabular}}
\newcommand{\et}{\end{tabular}}
\newcommand{\btab}{\begin{table}}
\newcommand{\etab}{\end{table}}
\newcommand{\bfig}{\begin{figure}}
\newcommand{\efig}{\end{figure}}
\newcommand{\bc}{\begin{center}}
\newcommand{\ec}{\end{center}}
\newcommand{\bit}{\begin{itemize}}
\newcommand{\eit}{\end{itemize}}
\newcommand{\nn}{\nonumber}

\newcommand{\tw}{\textwidth}
\newcommand{\ig}[1]{\includegraphics[width=#1\tw]}

\newcommand{\ssp}{\vspace{0.02\tw}}

\newcommand{\arr}{$\rightarrow$ }

\begin{document}


\title{Automation on the generation of genome scale metabolic models}

\author[a]{R.~Reyes}
\author[b,c]{D.~Gamermann\thanks{daniel.gamermann@ucv.es}}
\author[c]{A.~Montagud}
\author[c]{D.~Fuente}
\author[a]{J.~Triana}
\author[c]{J.~F.~Urchuegu\'ia}
\author[c]{P.~Fern\'andez de C\'ordoba}

\affil[a]{Universidad Pinar del R\'io ``Hermanos Sa\'iz Montes de Oca'',\\ Mart\'i 270, 20100, Pinar del R\'io, Cuba.}
\affil[b]{C\'atedra Energesis de Tecnolog\'ia Interdisciplinar, Universidad Cat\'olica de Valencia San Vicente M\'artir, \\ Guillem de Castro 94, E-46003, Valencia, Spain.}
\affil[c]{Instituto Universitario de Matem\'atica Pura y Aplicada, Universidad Polit\'ecnica de Valencia,\\  Camino de Vera 14, 46022 Valencia, Spain.}

\maketitle

\begin{abstract}
{\bf Background:} Nowadays, the reconstruction of genome scale metabolic models is a non-automatized and interactive process based on decision taking. This lengthy process usually requires a full year of one person's work in order to satisfactory collect, analyze and validate the list of all metabolic reactions present in a specific organism. In order to write this list, one manually has to go through a huge amount of genomic, metabolomic and physiological information. Currently, there is no optimal algorithm that allows one to automatically go through all this information and generate the models taking into account probabilistic criteria of unicity and completeness that a biologist would consider.

{\bf Results:} This work presents the automation of a methodology for the reconstruction of genome scale metabolic models for any organism. The methodology that follows is the automatized version of the steps implemented manually for the reconstruction of the genome scale metabolic model of a photosynthetic organism, {\it Synechocystis sp. PCC6803}. The steps for the reconstruction are implemented in a computational platform (COPABI) that generates the models from the probabilistic algorithms that have been developed.

{\bf Conclusions:} For validation of the developed algorithm robustness, the metabolic models of several organisms generated by the platform have been studied together with published models that have been manually curated. Network properties of the models like connectivity and average shortest mean path of the different models have been compared and analyzed.
\end{abstract}




\section{Introduction}

Since the second half of the twentieth century, the development of molecular biology has allowed fast advances in the understanding of the functions and working principles of cells and unicellular organisms at molecular level. In particular, high throughput experimental techniques of sequencing and analysis of genomic and proteomic information has given birth to rich web based databases of biological information on thousands of organisms, from prokaryotic bacteria to complex organisms like birds and mammals.

One of the new research fields that emerges from this panorama is system biology \cite{first}: the bottom-up approach in order to quantitatively explain the properties of biological systems from the modeling and simulation of the interactions and characteristics of its macromolecular components. From these systematic studies based on mathematical modeling and computational simulations, yet a new discipline appears: synthetic biology, which is focused on the design and construction of {\it ``a la carte''} new biological entities with new biological functions \cite{heinemann}. This new field of biotechnology lies in the limit between biology and engineering and aims at the partial design of modified organisms for different technological applications. For this purpose rational design principles of engineering must be combined with the available biological information and biotechnological techniques. The huge amount of biological information and the complexity of the engineering principles and analysis tools make it evident the need for a good computational platform that aids in the data mining and design development for artificial biological systems.

One of the corner stones of systems biology is the reconstruction of genome scale metabolic models. This means to gather information on all enzymatic reactions that take place in an organism based on genetic information available for this organism's genome. The current state of art of this process requires the effort of a specialist during a long term period (usually one year) in order to collect the available information from many different databases and the literature. Currently there are few software applications specially designed to help in this specific task. Moreover, different available software applications are too specific for a few determined tasks that do not embrace all steps in the whole process, leaving huge gaps that must be filled manually by the researcher.

The genome-scale metabolic reconstruction is the starting point of many different researches and applications, like the determination of the metabolic capacities or the determination of protocols for an optimal growing strategy for some organism and in particular, the search for potential sites for metabolic engineering \cite{oberhardt}. The aim of metabolic engineering is the modification and/or introduction of biochemical reactions with technologies like recombining DNA, in order to optimize the production of some metabolites of interest, to redirect metabolic fluxes to new pathways or even to extend the metabolic capacities of an organism for the production of new metabolites. The accomplishment of these objectives for a specific organism depends on a good reconstruction of its metabolism, from which one can study the structure of the metabolic network and the consequences of adding or deleting specific genes \cite{lopez}.

In this context, several projects have been developed for the reconstruction of genome scale metabolic models with different ends, like the production of fuel from cyanobacteria \cite{montagud} or yeast.
~Other examples are the genome-scale metabolic reconstruction of the {\it Burkhoderia cenocepacia J2315} \cite{fang} for research of treatment in patients suffering from cystic fibrosis; the {\it Rhodobacter sphaeroides} \cite{rhodo}, capable of producing hydrogen, polyhydroxybutyrate and other biofuels; the {\it Clostridium beijerinckii} \cite{clostridium} capable of producing butanol.

Genome-scale metabolic models result from the integration of genomic, proteomic and metabolomic information obtained at different experimental levels. The study of these theoretical reconstructions of cell metabolism allows researchers to investigate emergent phenomena in biology, like the feedback control loops that regulate the organisms and other aspects of metabolic and genetic transcription and regulation. The metabolic models constitute an important tool for the comprehension of an organism, its metabolic capacities and prediction of the responses of it to different environmental and genetic changes. Moreover, they facilitate the development of strategies for the engineering of metabolic systems focused in an improvement of the metabolic efficiency.

The process in order to generate the models consists in a first step to collect all available information on the metabolome of a species, as well as all genes that code for the different enzymes that catalyze each one of the metabolic reactions that take place in cell metabolism. Other aspects that must be taken into account are the coenzymes and cofactors needed for the enzymatic catalysis, the stoichiometry and reversibility of the reactions, information on the biomass composition and metabolic regulation \cite{forster}. Among all possible applications of a metabolic model, there is the possibility of evaluating projects for production and optimization of a metabolite of interest \cite{oberhardt}. If a model is satisfactorily constructed, it should allow a realistic simulation of the organism's metabolism, submitted to different environmental and genetic perturbations. This simulation would represent, with its natural restrictions, a virtual organism or an {\it ''in silico''} cell in which one can apply different computational algorithms to explore possible flux distributions inside the cell subjected to different environmental conditions and genetic configurations \cite{montagud}. For the analysis of metabolic models there are several computational tools and algorithms already developed \cite{edwards}. Those include Flux Balance Analysis (FBA) \cite{varma,edwards}, Minimization of Metabolic Adjustments (MOMA) \cite{segre} and Metabolic Flux Analysis (MFA) \cite{schilling,varma2} among others.

Currently, researchers that work in the genome-scale metabolic reconstruction use different computational tools in order to accomplish different tasks. For example, in order to generate the list of biochemical reactions for a specific organism, the software Pathway Tools \cite{karp} is a common choice. This software allows the compilation of specific databases for proteins, enzymes and metabolites of a determined organism, and to obtain a rough draft for the network. Nevertheless, it might be said that this software obtains the list of reactions with no regard to associations between metabolic routes and the reactions, or to completeness and unicity criteria for such model. These drawbacks leave a considerable amount of work to be manually done by the biologist that has to consider the reversibility of all reactions, and the analysis of possible repeated reactions inside the model, as well as the inclusion of reactions studied for other organisms that complete specific metabolic pathways, but do not show up in the database due to gaps in the genomic annotation for the organism in hand. Other specific software might be found with different computational algorithms implemented, like the Optgene software that includes an evolutionary programming based method to rapidly identify gene deletion strategies for optimization of a desired phenotypic objective function \cite{patil2}. 

The lengthy work of reconstructing a genome-scale metabolic network and analyzing it would be much faster and greatly simplified if one could find all algorithms and computational tools needed in the same software or platform. The aim of this work is to present a platform developed by our research group in order to automatically generate genome-scale metabolic models. With respect to other software that might be found for this purpose, our platform has the advantage that it automatically takes into account the criteria for unicity of the biochemical reactions, and presents the possibility for the biologist to automatically complete gapped metabolic pathways based in probabilistic criteria and comparison of the same metabolic route in different organisms. Moreover, the platform produces the metabolic generated network in different outputs: either as a SBML file or directly as an OptGene file format that might be directly piped in other analysis software. For the generation of the OptGene file, the biologist is also given the possibility to choose the biomass composition among the metabolites appearing in the model and directly fulfill the restrictions for the flux analysis and balance.

The work is divided as follows: in the next section we explain the algorithm for the generation of genome-scale metabolic models implemented in a web based platform. The algorithm automatically obtains information from the KEGG database for a specific organism and constructs from it the list of reactions in its metabolism. Criteria of unicity and completeness are taken into account in order to cope with different enzymes catalyzing the same reactions or to fill in missing reactions. Section 3 presents an analysis done with the models generated by the platform. The generated OptGene files are used as inputs in different algorithms to study network properties of the reconstructed metabolic models and compare them with manually reconstructed models taken from the literature. In the end we present our conclusions, an overview and future perspectives.

\section{Algorithm Description}

The algorithm for automatically generating metabolic models comprises several steps: the information compilation from free access biological databases, following some interaction of the user with the platform in order to properly select the parameters for the probabilistic criteria and choices for the biomass components and restrictions, and finally application of unicity and completeness criteria and production of the output.

\subsection{Obtaining the Biological Information}

In the last years, applications of biotechnology in different areas of science and technology have considerably increased causing an exponential growth on the available information on different organisms about their genetics, regulation processes and metabolism. Such information, obtained by different techniques with growing efficiency, becomes part of huge databases, many of which are of free access. This information together with the vast published scientific works put in the hands of the researchers a rich ever-growing amount of data and information.

A start point for the genome-scale metabolic reconstruction is to obtain the relevant information about the organism for which the model is going to be generated, namely the list of reactions, genes, metabolites and enzymes present in the studied cell. This information is available from public free access databases like Biocyc \cite{karp2}, Kegg \cite{kanehisa}, Brenda \cite{chang}, Uniprot \cite{uniprot}, etc. Nevertheless, the lack of quality in some entries of the databases are an inconvenient that one must amend: false positives, false negatives, as well as objects wrongly annotated may pose obstacles in the efforts to compile a meaningful correct list of reactions \cite{weise}. As a consequence, the reconstruction must be done under strict control of all and each one of the reactions, the biomass equation must be based in constituent molecules and coherence and integrity of the network must be prerequisites for the generation of a quality and useful model \cite{feist}.

The first objective of the project is to obtain the relevant rough information. For this aim we have chosen to use Kegg API web service offered by the Kegg database. This service allows one to access the Kegg system via SOAP/WSDL which offers valuable tools in order to access the available information in the mentioned database. These tools are designed for the search of cellular biochemical processes as well as to analyze the universe of genes and completed genomic sequences of thousands of organisms. The users might access Kegg API server by the SOAP technology through HTTP protocol.

The Simple Object Access Protocol (SOAP) defines how two objects in different processes may communicate trough exchange of XML (eXtensible Markup Language) data for a variety of bioinformatic applications. With this protocol, an application running in a machine anywhere in the world can use algorithms, data and resources stored in different servers \cite{snell}. Web Services Description Language (WSDL) is based in XML and allows one to have the description of a web service, specifying the abstract interface trough which a client may access the service and the details on how to use it \cite{url1}.

These technologies allowed us to access Kegg API and to construct a service web client using Java \cite{url2} as programming language and Netbeans \cite{url3} 6.8 as integrated development environment. This way we obtained biological information from the definition of directional data model that relates biological elements for its storage in a database constructed in Postgres \cite{saez,bourne,altman}, taking into account the relationships among them and their importance in the reconstruction of the genome-scale metabolic models. The relevant biological elements identified at this point are the organism, gene, pathway, enzyme, reaction, compound (metabolites and glycans) and the references. In figure \ref{fig:kegg} we show an schema on how information is organized and related in the KEGG data base, and the methods to obtain the relations between the different elements.

\bfig
\bc
\ig{1.0}{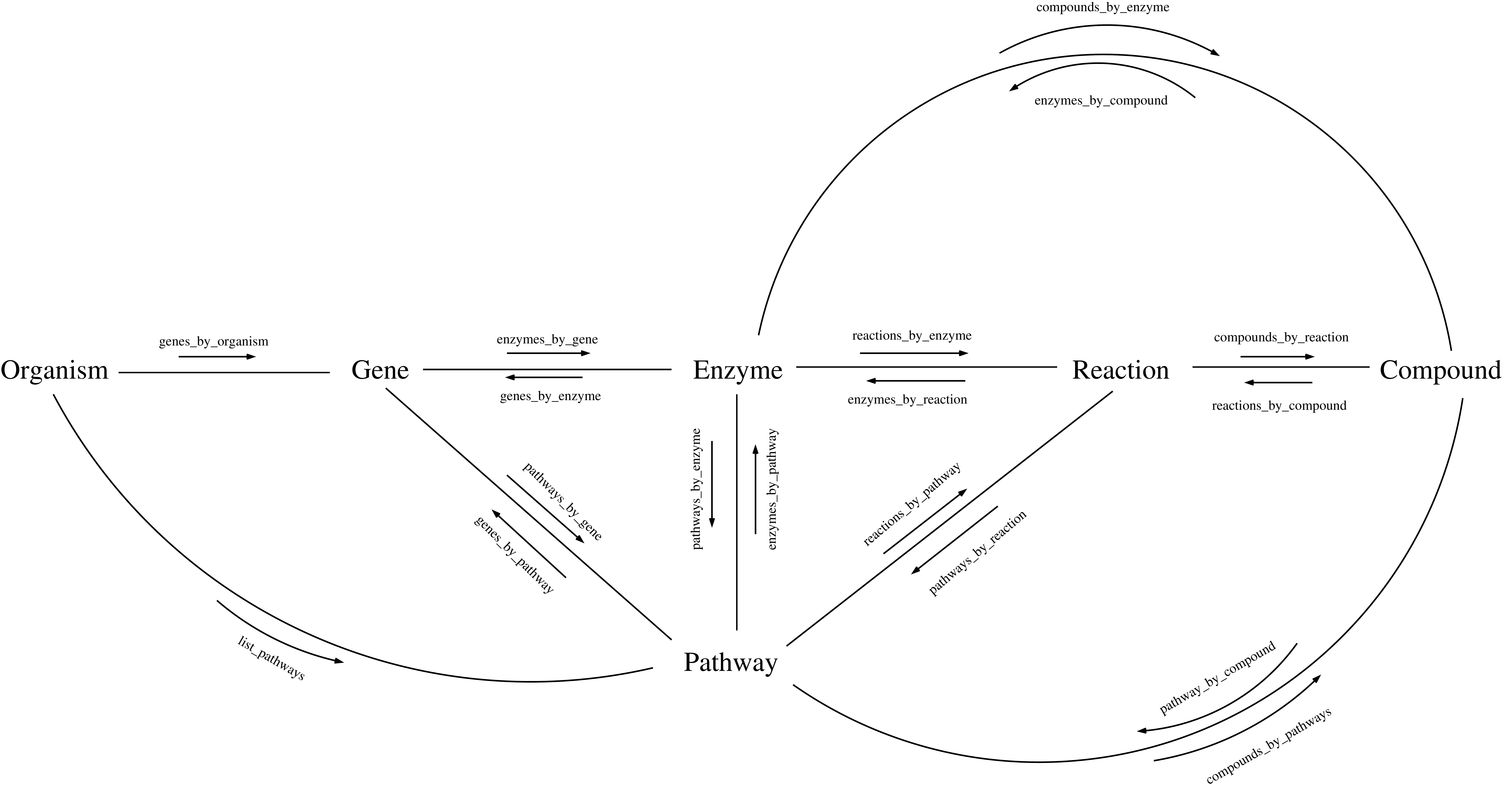}
\ec
\caption{Scheme of KEGG information with the WSDL methods for obtaining the information. The same methods that appear referencing compounds also exist for glycans (which are not shown in the figure).}\label{fig:kegg}
\efig


\subsection{Computational Platform for the Access of Biological Information (COPABI)}

All the biological databases available in the internet offer a web application to show their information, as well as methods for exporting this information in different forms. The first aim of our project is to implement an application that allows biologists to consult the relevant information from the database and to automatically generate metabolic models from it, where probabilist criteria for unicity and completeness are taken into account in order to generate more accurate models with greater quality in an efficient way. The platform resulting from the project is named COPABI from COmputational Platform for the Access of Biological Information.

Every web application requires for its publication a web server which is the responsible for waiting client requests and to answer them. In the case of COPABI the chosen web server application is Apache, an open source code that works under any platform that has become one of the best web servers in terms of efficiency, functionality and speed \cite{url4}.

In order to select the tools for the implementation of COPABI, it has been taken into account that among the possible web applications there are two major groups: the languages that run from the side of the server, like PHP and the ones that run from the side of the client, like HTML \cite{url5}, Javascript \cite{Eguiluz}, etc. In the case of COPABI, PHP has been chosen for this is an open source and very popular programming language, especially in web development \cite{url6,age}. As web application framework (WAF) we have used Codeigniter 1.7.3 \cite{url7}. The aim of this framework is to aid in the development of code, from a rich sample of libraries, a simple interface and a logical structure for accessing the libraries. Javascript has been used for interactivity of the pages and for validations and the completion of data in the search for biological information.

\subsection{Formats for the Output}

Different analysis tools in system biology make use of different file formats. Therefore, a good choice for the output file format will be determinant in the uses and utility of the generated models. Some examples of file formats used by different databases are: SBML \cite{hucka} (System Biology Markup Language), FASTA \cite{url8} (Fast All), BLAST \cite{url9} (Basic Local Alignment Search Tool) and Kegg presents its information in a particular markup language called KGML \cite{kegg}.

The lack of a standard makes more difficult the use of different softwares that usually have specific formats for their inputs and output. Sometimes the user has to go through the lengthy work of converting the information from one format to the other. One of the most versatile formats is certainly SBML, since it is a description language based on XML that can be used to represent models of different biological processes like metabolic network, cellular signaling pathways, genetic regulation networks, among others. A very useful file format specific for metabolic networks is the one following the input requisites for the OptGene software \cite{patil2}, also called BioOpt and used in BioMet toolbox (www.sysbio.se/biomet).

For the COPABI platform these two output file formats have been chosen: the standard SBML level 2 version 1 and the OptGene file format, which can be readily used for flux balance analysis.

\subsection{Construction of the Models}

For the generation of metabolic models for any organism, the applied methodology follows the same steps used in the manual reconstruction of the first model of a photosynthetic organism, the {\it Synechocystis sp.} PCC6803 \cite{montagud} also taking into account the probabilist criteria followed in this paper in order to deal with duplicated and missing reactions, which will be discussed later.

About the list of reactions in the reconstruction of the metabolic model, before applying the probabilist criteria one has to properly organize and identify the chemical reactions. Three issues should be noted here:

\bit
\item First is the compilation of all chemical reactions of a particular pathway present in the organism. There are two kinds of reactions that one has to take into account: most reactions are catalyzed by enzymes and each enzyme has a code called EC (Enzyme Commission), each reaction in the model receives this number as its identifier. On the other hand, there are a few reactions in some pathways that are not catalyzed by enzymes, they are spontaneous reactions, instead. These spontaneous chemical reactions receive as their identifier the name ``non-enzymatic'' and a number after it indicating the order in which these reactions appeared.
\item Next is the related to enzymes that can catalyze different reactions of the same type. This usually happens because different substrates have very similar structure and can couple to the enzyme which performs its catalytic activity. In these cases, next to the EC number of the enzyme, in the identifier comes an extra letter (a,b,c,...).
\item In a last step the reversibility (or irreversibility) of each reaction in a pathway is checked from kgml files for every pathway present in an organism.
\eit

Example of output from {\it Synechocystis sp.} PCC6803 in OptGene format (the symbols \# indicate comments and we use it for specifying the name of the metabolic pathway):

\ssp

-REACTIONS

\# Glycolysis / Gluconeogenesis 

1.2.4.1a:   Pyruvate +  Thiamin diphosphate -$>$  CO2 +  
~~~~2-(alpha-Hydroxyethyl)thiamine diphosphate 

2.7.1.40a:   ATP +  Pyruvate -$>$  ADP +  Phosphoenolpyruvate 

6.2.1.1a:   ATP +  CoA +  Acetate -$>$  Diphosphate +  AMP +  Acetyl-CoA 

1.2.1.5a:   H2O +  NAD+ +  Acetaldehyde $<$-$>$  NADH +  Acetate +  H+ 

1.1.1.2:  NADP+ +  Ethanol $<$-$>$  NADPH +  H+ +  Acetaldehyde

\# Citrate cycle (TCA cycle) 

1.1.1.42a:   Oxalosuccinate $<$-$>$  CO2 +  2-Oxoglutarate 

1.1.1.37:  NAD+ +  (S)-Malate -$>$  NADH +  Oxaloacetate +  H+ 

2.3.3.1:  CoA +  Citrate $<$-$>$  H2O +  Acetyl-CoA +  Oxaloacetate 

6.2.1.5a:   ATP +  CoA +  Succinate $<$-$>$  ADP +  Orthophosphate +  Succinyl-CoA

\subsection{Probabilistic Criteria}

Until this point, the reconstructed model will have exactly the same information stored in the database. The next steps are the implementation of automatic algorithms that will take into account the probabilistic criteria in order to complete missing gaps in some metabolic pathways (completeness) or to exclude duplicated reactions (unicity).

\subsubsection{Unicity}

For the unicity criteria the algorithm identifies reactions that appear more than once and identifies their enzymes. Repeated reactions must be eliminated, the criteria to choose which reaction is eliminated is the following: the enzyme that appears less frequently in the model is not eliminated. As an example we show the reactions:

\ssp

1.2.1.12a: A + B $<$-$>$ C + D

1.2.1.12b: G + E $<$-$>$ K + L

1.2.1.12c: P + V -$>$ Y

1.1.1.1: G + E $<$-$>$ K + L

\ssp

In the example enzyme EC1.2.1.12 catalyzes 3 different reactions, one of which is the same as the reaction catalyzed by enzyme EC1.1.1.1. Following the algorithm criteria, reaction 1.2.1.12b is eliminated from the metabolic model.

\subsubsection{Completeness}

The addition of new reactions to a metabolic model is associated to the comparison between the information available about the metabolic reactions in a determined pathway in a specific organism with a general pathway (theoretical one) generated from the compilation of all metabolic reactions present in ``all'' organisms in nature. In many cases the genomes are not perfectly annotated and some genes are missing from the annotation. This comparison of pathways is a tool to help biologists identify these missing genes. 

The reactions associated to gaps identified in a particular pathway are going to be added to the metabolic model if they satisfy the following criteria:

\bit
\item The reactions have as final product a metabolite belonging to the biomass equation.
\item The reactions present in the model in this pathway correspond to, at least, a determined percentage of the whole general pathway.
\eit

The COPABI interface allows the user to chose the metabolites that compose the biomass equation, as well as the percentage value used in the second criteria. 

The reactions added to the model with these criteria are added to the end of the output after a comment (\# not pres ! $\sim\sim$IMPORTANT, following X reactions not in sequence!!$\sim\sim$) and their identifier is the correspondent EC number with the symbol ``$\cdot$'' preceding it.

\section{Results - Validation of the Models}

For the validation of the metabolic models generated by the COPABI platform, we have analyzed the models generated for determined organisms with metabolic models manually curated taken from the literature. In a first step we analyze general properties of the generated models (number of metabolites, reactions, ...) and properties of the networks described by the metabolic model.

From the network point of view each metabolite of a model can be thought as a node and each reaction represents links between the metabolites in the lefthand-side with the metabolites on the righthand-side of the reaction equation. These links can be directed if one takes into account the direction of each reaction and the reversibility of the reactions or undirected if one neglects this information.

In a first step of the analysis, our algorithms make an automatic debug of the metabolic models. This means look for bad reactions, those being reactions without substrates or products (some transport reactions in the sbml files taken from the literature present this issue), reactions where the same metabolite appears as substrate and product or reactions decoupled from the network, meaning that at least one substrate and one product of the reaction appear only in this reaction and nowhere else. All these bad reactions are excluded from the models before any calculation is done, because these bad reactions add up errors and uncertainties to the results.

Each metabolic model now represents a network and, in a first step, we have chosen to work with the directed version of it, meaning that the links connecting two metabolites have direction from the substrate to the product and in reversible reactions the pair of metabolites would have two links of opposite directions connecting them.

For each organism two versions of the metabolic model have been generated with two different values for the parameter appearing in the decision taking process for the completeness criteria, explained in the section before. In one model, the parameter is chosen to be 100\% and in the other 10\%, two possible extremes. Taking this parameter to be 100\% means that only reactions for enzymes that are annotated in the genome of an organism will be written to the model. On the other hand if the parameter is taken to be 10\% (a fare low value) means that if one out of ten reactions in a pathway are annotated in the genome, the generated model will have all reactions in this pathway. In the end, for each organism there are three models being studied, two automatically generated by the COPABI platform and one manually constructed taken from published works. The models taken from the literature correspond to the following organisms: the {\it Synechocystis sp PCC6803} \cite{montagud}, {\it Synechococcus elongatus PCC7942} \cite{julian}, {\it Burkhoderia cenocepacia J2315} \cite{fang}, {\it Rhodobacter sphaeroides} \cite{rhodo}, {\it Clostridium beijerinckii} \cite{clostridium}, {\it Mycoplasma genitalium} \cite{mge}, {\it Lactobacillus plantarum} \cite{lpl}, {\it Thermotoga maritima} \cite{tma} and {\it Yerisinia pestis} \cite{ypk}. 

In table \ref{tab:comparison} we show results for the general comparison of the models. Below we explain each column of the table separately:

\bit
\item \# Met. \arr The number of different metabolites (or compounds) found in the model.
\item \# Reac. \arr The number of reactions present in the model (after excluding bad reactions).
\item \% Rev \arr The percentage of the reactions that are reversible.
\item \% Irr \arr The percentage of the reactions that are irreversible.
\item ASP \arr The average shortest path. For each pair of metabolites in the model we have used Dijkstra's algorithm to calculate the shortest path connecting them in the network. For all pair of metabolites where the shortest path was found, the average value was calculated (pair of metabolites not connected by any path were left out).
\item $\sigma_{\textrm{ASP}}$ \arr The standard deviation for the ASP calculation.
\item $N_R$ \arr The number of pair of metabolites for which a path connecting them was found.
\item $N_U$ \arr The number of pair of metabolites for which a path connecting them was not found. One should note that the network is directed, so, metabolites that have no link pointing in their direction can not be reached by any pair and are, therefore, either external metabolites that should be absorbed by the cell from the environment or are badly incorporated to the model.
\eit

\btab
\caption{General Comparison. The explanation of each column is in the text.}\label{tab:comparison}
\bc
\bt{c|cccccccc}
Org. & \# Met. & \# Reac. & \% Rev. & \% Irr. & ASP & $\sigma_{\textrm{ASP}}$ & $N_R$ & $N_U$ \\
\hline
\hline
syn\_lit & 803 & 893 & 34.49 & 65.51 & 3.51 & 1.15 & 494446 & 150363 \\
syn\_10 & 707 & 718 & 37.74 & 62.26 & 3.2 & 0.90 & 355430 & 144419 \\
syn\_100 & 656 & 640 & 36.40 & 63.60 & 3.29 & 0.94 & 295093 & 135243 \\
\hline
syf\_lit & 777 & 847 & 36.01 & 63.99 & 3.55 & 1.19 & 475612 & 128117 \\
syf\_10 & 711 & 705 & 37.02 & 62.98 & 3.19 & 0.88 & 356066 & 149455 \\
syf\_100 & 655 & 622 & 35.05 & 64.95 & 3.32 & 0.95 & 292390 & 136635 \\
\hline
cbe\_lit & 732 & 856 & 27.22 & 72.78 & 3.05 & 0.82 & 409910 & 125914 \\
cbe\_10 & 752 & 808 & 40.22 & 59.78 & 3.21 & 0.88 & 412228 & 153276 \\
cbe\_100 & 693 & 733 & 38.2 & 61.8 & 3.33 & 0.97 & 335276 & 144973 \\
\hline
tma\_lit & 583 & 612 & 41.67 & 58.33 & 3.19 & 0.96 & 242290 & 97599 \\
tma\_10 & 566 & 614 & 46.09 & 53.91 & 3.06 & 0.83 & 250504 & 69852 \\
tma\_100 & 489 & 517 & 44.1 & 55.9 & 3.24 & 0.91 & 183170 & 55951 \\
\hline
bcj\_lit & 792 & 847 & 27.63 & 72.37 & 3.04 & 0.83 & 523487 & 103777 \\
bcj\_10 & 955 & 1018 & 37.03 & 62.97 & 3.25 & 0.88 & 632355 & 279670 \\
bcj\_100 & 907 & 948 & 36.29 & 63.71 & 3.32 & 0.92 & 564967 & 257682 \\
\hline
mge\_lit & 342 & 262 & 40.08 & 59.92 & 3.00 & 0.99 & 83279 & 33685 \\
mge\_10 & 268 & 254 & 48.82 & 51.18 & 2.89 & 0.82 & 54311 & 17513 \\
mge\_100 & 116 & 104 & 55.77 & 44.23 & 3.42 & 1.25 & 11543 & 1913 \\
\hline
eco\_lit & 1034 & 1435 & 14.91 & 85.09 & 3.19 & 0.90 & 912770 & 156386 \\
eco\_10 & 888 & 1017 & 39.53 & 60.47 & 3.19 & 0.84 & 576962 & 211582 \\
eco\_100 & 846 & 968 & 38.95 & 61.05 & 3.26 & 0.87 & 525007 & 190709 \\
\hline
lpl\_lit & 513 & 526 & 31.37 & 68.63 & 2.97 & 0.82 & 221786 & 41383 \\
lpl\_10 & 566 & 595 & 41.85 & 58.15 & 3.14 & 0.83 & 233640 & 86716 \\
lpl\_100 & 492 & 512 & 41.8 & 58.2 & 3.23 & 0.88 & 173827 & 68237 \\
\hline
rsp\_lit & 788 & 863 & 64.31 & 35.69 & 2.74 & 0.68 & 593663 & 27281 \\
rsp\_10 & 869 & 934 & 41.65 & 58.35 & 3.16 & 0.82 & 543333 & 211828 \\
rsp\_100 & 827 & 873 & 40.21 & 59.79 & 3.24 & 0.87 & 485209 & 198720 \\
\hline
ypk\_lit & 817 & 948 & 29.85 & 70.15 & 3.04 & 0.85 & 339398 & 142238 \\
ypk\_10 & 838 & 945 & 39.47 & 60.53 & 3.18 & 0.84 & 520075 & 182169 \\
ypk\_100 & 779 & 891 & 39.62 & 60.38 & 3.25 & 0.89 & 444404 & 162437 \\
\et
\ec
\etab

As one can see from the table, although the networks usually have hundreds of different metabolites, two different metabolites are on average only three steps apart from each other. As a consequence, the whole network should very fast respond to changes in any of the metabolites' concentrations or to environmental perturbations. This closeness of the nodes in the network is known as small world behavior and is a consequence of a property of the network connectivity know as scale free distribution. Metabolic networks are known to follow a free scale distribution for node connectivity, meaning that the number of nodes ($P$) with some number of connections ($x$) follows a power law: $P(x)\sim x^{-\gamma}$ where $\gamma$ is usually a number between 2 and 3. From this law one concludes that there are very few nodes with a large number of connections (these are called hubs) and most of the nodes have very few connections.

For studying node connectivity, for each metabolite an algorithm counts in how many reactions it appears as a substrate (or product in reversible reactions). In figure \ref{fig:connect} we show results for the three metabolic models of some organisms. It is clear from these plots the tendency of the distribution to follow a power law.

\bfig
\bc
\bt{cc}
\ig{0.5}{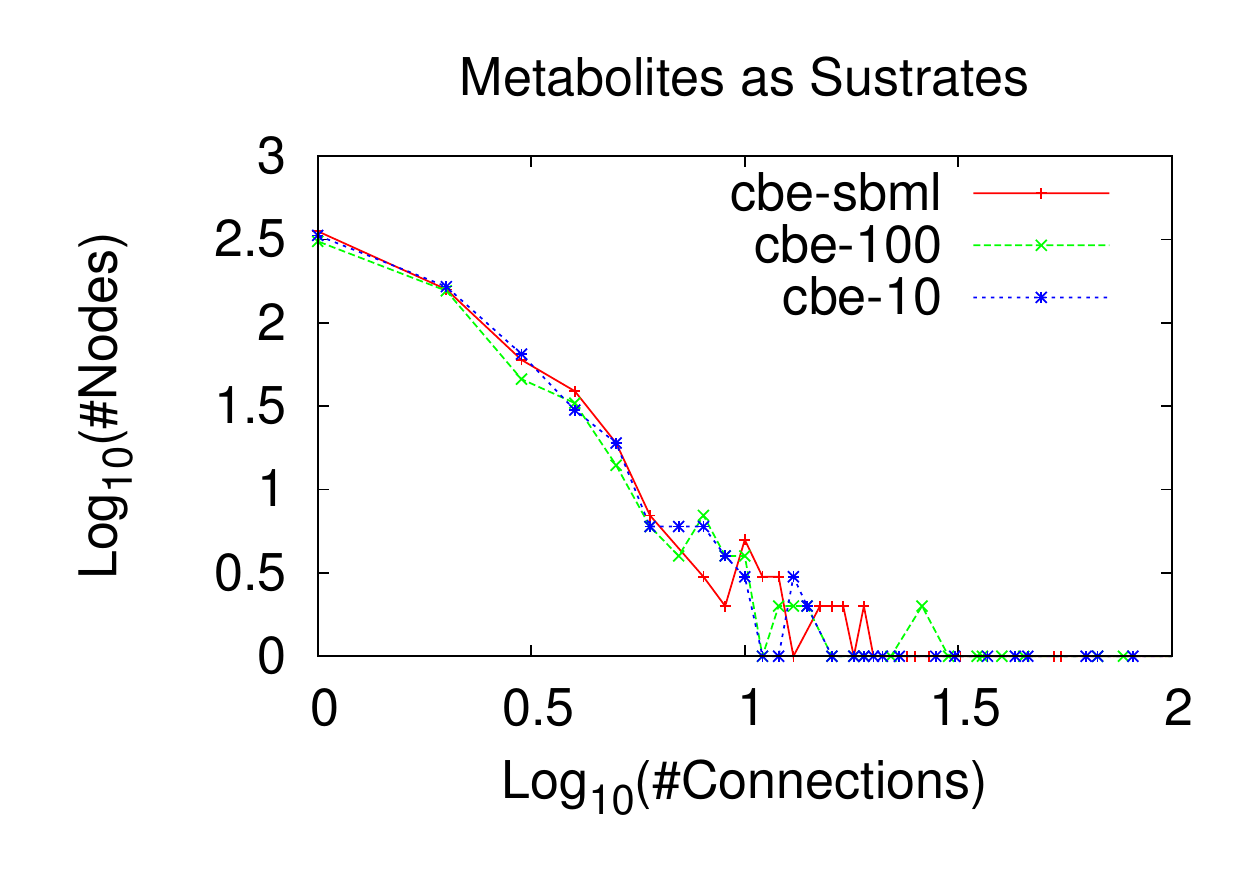} & \ig{0.5}{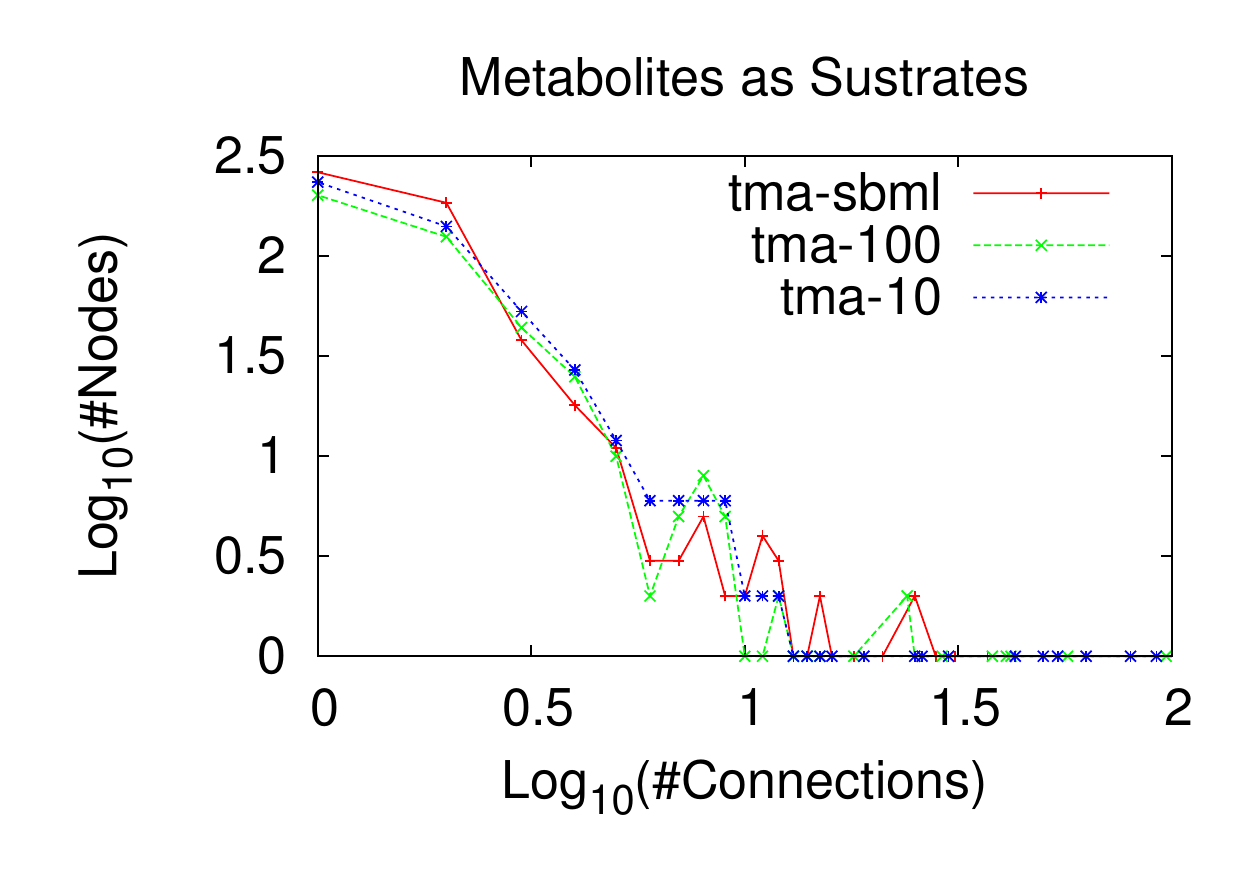}  \\
\ig{0.5}{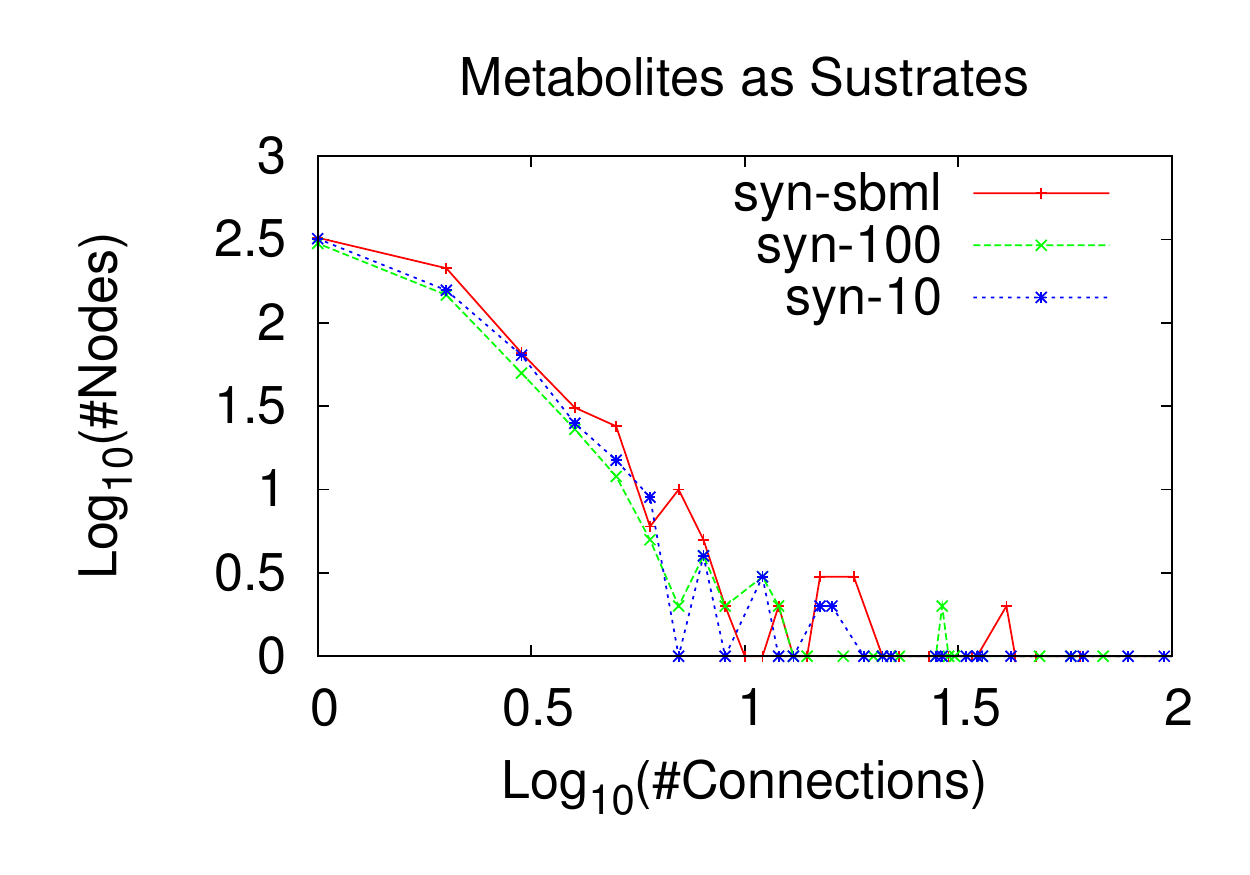} & \ig{0.5}{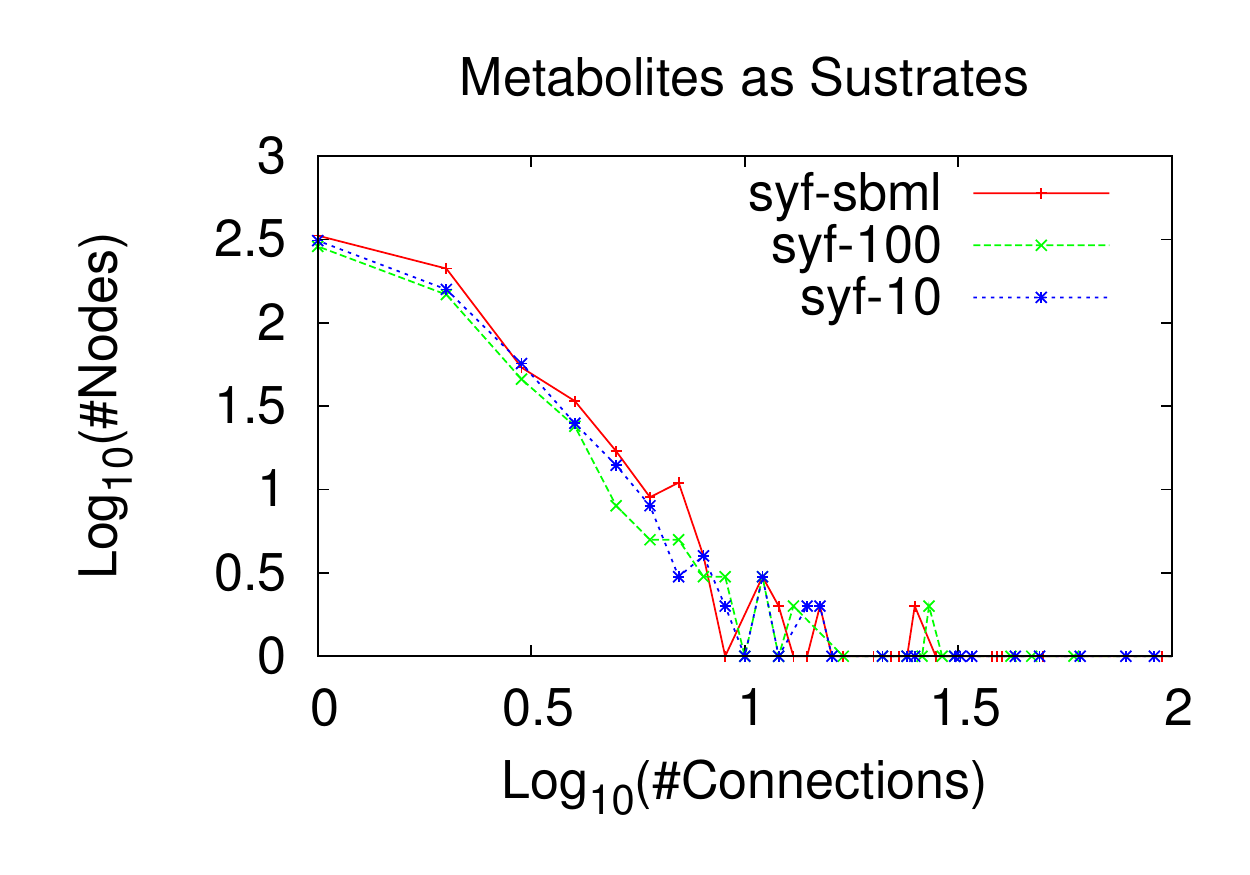}  \\
\et
\ec
\caption{Connectivity distribution for the three metabolic models in different organisms.}\label{fig:connect}
\efig

As one can see, all metabolic networks are very similar when their global network properties are studied. Therefore, in order to be able to differentiate the metabolic network of one organism from the network of another different organism, one has to look into the details of the networks, namely specific metabolites and hubs that are particular for each organism.

In order to find these differences we are going to define a similarity parameter in order to measure the degree of similarity of two metabolic networks. Two criteria are taken into account in the definition of this parameter, first the metabolites present in each metabolic network and the degree (number of connections) of each metabolite to all others. Since important metabolites for one organism, might be different from the essential metabolites in another one, we would like to take into account in the comparison also the identity of the metabolites in the different connections and not only the number of connections. This step is rather tricky because the metabolites names used in the metabolic models taken from literature do not follow any standard and the authors of each model chose different abbreviations and names for each compound. For some models, however, the authors have also made available the association of each compound name used in their models with a Kegg identifier. For these models we were able to construct an algorithm that translated the metabolic model to the same standard metabolite names used by kegg and therefore we were able to also compare the metabolite identities.

\subsection{Network Comparison}

Our goal here is to define a distance between two networks. We call it a distance in the sense that the bigger the value of this number, the more different the two networks will be, while the smaller the number is, the closer the networks are. 

Given two metabolic networks, each one has a set of metabolite (let's call the set in each network by set $A$ and set $B$). Among all metabolites in the two networks, there are three different set: metabolites particular to network $A$, metabolites particular to network $B$ and metabolites common to the two networks:

\be
A\cup B&=& \underbrace{(A\cap \bar{B})}_{\textrm{only in A}}  \cup  \underbrace{(A\cap B)}_{\textrm{Common}} \cup \underbrace{(\bar{A}\cap B)}_{\textrm{Only in B}}
\ee

Now, let's consider the connections of the metabolites: Each metabolite $i$ has $n_i$ connections in total and $n_{\alpha i}$ connections only to metabolites in the set $A\cap \bar{B}$, $n_{\beta i}$ connections only to metabolites in the set $B\cap \bar{A}$ and $n_{\gamma i}$ connections only to metabolites in the set $A\cap B$.

Let's define the number of metabolites in each set and the total number of connections inside each set:

\be
N_\alpha &=& \parallel A\cap \bar{B} \parallel \\
N_\beta &=& \parallel B\cap \bar{A} \parallel \\
N_\gamma &=& \parallel A\cap B \parallel  \\
N_A &=& \sum_{i\in A\cap \bar{B}} n_i  \\
N_B &=& \sum_{i\in B\cap \bar{A}} n_i  \\
N_C &=& \sum_{i\in A\cap B} n_i 
\ee
Here, $\parallel C \parallel$ means the number of elements in the set $C$.

Now, for each set, lets sum the proportion of connections of each metabolite to metabolites inside the set, weighted by the inverse of the total number of connections and averaged for all metabolites. 

\be
p_{A i}&=& \frac{n_{\alpha i}}{n_i} \nn \\
\alpha &=& \frac{N_A}{N_\alpha}\sum_{i\in A\cap \bar{B}}\frac{1}{n_i}p_{A i} \nn
\ee
and analogously we define $\beta$ and $\gamma$ for the metabolites in the other two sets.

The distance between the two networks is defined as:

\be
\textrm{dist}&=& \frac{\alpha+\beta}{2\gamma}\nn
\ee

For an identical network, $\alpha$ and $\beta$ are zero, so dist=0. For two networks which have not a single metabolite in common, $\gamma$=0 and so dist=$\infty$.

For validating our models we proceeded by doing this rough comparison, calculating the distance between each one of the automatically generated models to each model taken from the literature. In Table \ref{tab:compa} we show results for the comparison among 5 models taken from the literature and the automatically generated ones, one can observe that for each organism, the biggest value for the comparison in each column is when comparing the models for the same organism.

\btab[h]
\bc
\caption{Distance comparison between automatically generated and manually curated models. In bold face are the largest number in each row, showing that the best comparison between models is when comparing models for the same species.} \label{tab:compa}

\ssp

\bt{c|cccccc}
\hline
 org       & syn\_lit & syf\_lit & cbe\_lit  & rsp\_lit  & ypk\_lit & tma\_lit   \\ 
\hline
\hline
mge\_10     & 1.246      & 1.254      & 0.401      & 0.695      & 1.003      & 1.541   \\ 
\hline
lpl\_10     & 0.815      & 0.755      & 0.121      & 0.317      & 0.476      & 0.834    \\ 
\hline
syn\_10     & {\bf 0.47} & 0.527      & 0.183      & 0.248      & 0.626      & 1.065      \\ 
\hline
syf\_10     & 0.54       & {\bf 0.496}& 0.18       & 0.255      & 0.628      & 0.999    \\ 
\hline
cbe\_10     & 0.697      & 0.699      & {\bf 0.076}& 0.212      & 0.413      & 0.814    \\ 
\hline
bcj\_10     & 0.708      & 0.721      & 0.156      & 0.183      & 0.459      & 1.063      \\ 
\hline
eco\_10     & 0.748      & 0.799      & 0.13       & 0.204      & 0.387      & 0.959    \\ 
\hline
tma\_10     & 0.72       & 0.7        & 0.103      & 0.278      & 0.498      & {\bf 0.636}\\ 
\hline
rsp\_10     & 0.735      & 0.741      & 0.157      & {\bf 0.138}& 0.549      & 1.103      \\ 
\hline
ypk\_10     & 0.772      & 0.782      & 0.12       & 0.181      & {\bf 0.324}& 0.882      \\ 
\hline
\et
\ec
\etab

This is a rough comparison, since the identification of the metabolites' names is not perfect. Moreover, since the automatically generated models have not yet been used in flux balance analysis, there is no distinction between internal and external metabolites, and we used a version of these models where no biomass was defined. This introduces lots of errors and uncertainties in the comparison made, because the models from the literature do have these features defined and they appear, from the point of view of the comparing algorithm, as new and different metabolites for which there will be no counterpart in the automatically generated models. Despite these pitfalls, the features already contained in the models are enough to differentiate the different organisms when compared to models from the literature.

\section{Discussion}

We presented here the metabolic models automatically generated by an algorithm developed by our research group. This algorithm, implemented in the COPABI platform, is able to automatically download genomic, metabolomic and proteomic information from the kegg database and construct, from this information, a genome-scale metabolic model either in OptGene or xml (sbml) file formats. The tools presented here will be soon made available in the web.

The metabolic models generated have been throughly analyzed by standard algorithms in order to calculate average shortest mean path between nodes in the network and connectivity distribution. Commonly observed features in these networks are small world behavior and scale free distribution for the nodes degree.

After a general analysis of the global properties of the networks, we proceeded to compare the automatically generated models with model manually curated we took from published works. For this task a distance parameter between metabolic networks has been defined. The comparison shows that the automatically generated models are consistent with models constructed manually found in the literature.

The reconstruction of genome-scale metabolic models is an important step in different areas of research related to systems biology. Nowadays, this is a lengthy and slow process that might take over a year to be completed for a single organism. The algorithm developed here will certainly speed up the process and help researches to gain access to reconstructed models for any organism in just a few days period.


\section*{acknowledgements}

The authors would like to thank Alberto Conejero for useful discussions.
This work has been funded by MICINN TIN2009-12359 project ArtBioCom from the Spanish Ministerio de Educaci\'on y Ciencia.


\bibliographystyle{spbasic}

\end{document}